\documentclass[12pt]{iopart}

\usepackage{epsfig}

\begin{document}

\title{A comparative study for the pair-creation 
contact process using series expansions}

\author{C. E. Fiore, W. G. Dantas and M. J. de Oliveira}

\address{Instituto de F\'{\i}sica,
Universidade de S\~{a}o Paulo,
Caixa Postal 66318
05315-970 S\~{a}o Paulo, S\~{a}o Paulo, Brazil}

\begin{abstract}

A comparative study between two distinct 
perturbative series expansions for the pair-creation contact process
is presented.
In contrast to the ordinary contact process, 
whose supercritical series expansions
provide accurate estimates for its critical behavior,  
the supercritical approach does not work properly when applied to
the pair-creation process. 
To circumvent this problem 
a procedure is introduced in which one-site creation is added to
the pair-creation.
An alternative method  is the generation of 
subcritical series expansions which works even for the case
of the pure pair-creation process.  Differently from the 
supercritical case, the subcritical series yields estimates 
that are  compatible with numerical simulations. 

\end{abstract} 


\maketitle

\section{Introduction}

The critical properties of nonequilibrium systems have attracted  
attention and have seen a great
development in the last years \cite{md99,priv97,h06}.
This may be  credited to their 
resemblance with critical phenomena  
occurring in systems in thermodynamic equilibrium.  
In equilibrium as well as in nonequilibrium phase transitions
there is a singular dependence of the 
steady-state properties upon the control parameters, 
marking a transition between distinct regimes.  
Other common features include long-range
correlations, well-defined order parameter and 
singularities characterized by critical exponents that are
grouped in universality classes \cite{odor04}.

Due to absence of a general theory for the nonequilibrium
regime, these properties are analyzed in particular models, 
trying to form a complete picture. In particular, 
a class of simple models used to perform this task 
is constituted by lattice models with absorbing states.  
The presence of an absorbing state is
sufficient for breaking the detailed balance condition,
making these models intrinsically irreversible. 
In the steady-state regime, these models 
display a phase transition from an absorbing
state to an active state, as the control parameter is varied.

A very robust universality class for these models is  
the directed percolation (DP) class. It describes   
the critical properties of any system with a 
phase transition between an absorbing and an active state 
characterized by a scalar order parameter, short-range 
interactions and no conservation laws \cite{j81}.

Usually, the tool utilized in the study of these problems 
is the numerical simulation, but analytic approaches
are also used as an important tool in this 
task leading, in some cases, to very precise
results.  As in other fields of physics, 
simulational and analytical approaches are complementary
in the study of phase transitions in equilibrium and nonequilibrium 
systems \cite{h00}.
One of these analytical approaches is the series expansion \cite{dj91} 
that has provided the most precise estimates of the DP exponents in
one-dimension  \cite{j99}. Inspired by the  
equilibrium case, the series for nonequilibrium systems can be either
supercritical or subcritical, in analogy with the 
high- and low-temperature expansions in equilibrium systems.

A pioneer work that uses series expansions for lattice models 
with absorbing states was developed by Dickman and Jensen \cite{dj91}, 
in their study of the critical properties of the contact process (CP).  
The contact process is an icon among the models that display a 
phase transition between an absorbing and active state.  
Initially the CP was proposed as a model for spreading of an 
epidemic disease \cite{h74} and has become the 
``Ising model'' for the DP universality class. 
Using subcritical and supercritical
series, Dickman and Jensen obtained the critical point and its
associate exponents. Their results indicate that the supercritical case
works better than the subcritical series.

Recently, de Oliveira \cite{mj06} proposed an alternative approach 
for generating the subcritical expansion 
obtaining critical values comparable to those of the supercritical
series \cite{dj91}.  The difference between the approaches is 
that in \cite{mj06}, the  non-perturbated operator (associated with 
the  annihilation of particles) should be
diagonalized, whereas this procedure
is not required for the supercritical series.

In this paper, a comparative study 
between subcritical and supercritical series expansion is performed
for the pair-creation contact process (PCCP) \cite{dicktom91}.  
Although this model does not show any surprise concerning the 
critical properties, since it belongs to the DP universality
class as the ordinary CP, it is interesting for the comparative 
analysis.  We will show that the original 
approach for the supercritical series \cite{dj91}
does not work properly, presenting problems in its construction.  
Even an alternative version of the model, that would overcome these
problems, does not lead to precise results for the critical point.
On the other hand, the subcritical series gives results,
without any maneuver, that are comparable to those obtained by numerical
simulations \cite{mariofiore}. 

This paper is organized as follows. 
In section 2 we present the model and the operator formalism.
In section 3 we show the derivation of the 
supercritical series, discussing its problems.
In section 4, we attempt to overcome the problem by
introducing a modified PCCP and 
deriving the supercritical series.
In section 5 we present the approach for the subcritical 
case and results coming from the Pad\'e analysis for
both pure and modified PCCP.
Finally, in section 6 the conclusions and final discussions 
of this work may be found.       

\section{Pair-creation contact process and operator formalism}

We consider an interacting particle system on a one-dimensional
lattice with $L$ sites.
The system  evolves in the time according to a Markovian
process with local and irreversible rules.   
The configurations are described in terms of occupation
variables, $\eta_i$, with $\eta_i=0,1$ according to whether 
the site $i$ is empty
or occupied by a particle, respectively.
The time evolution of the probability $P(\eta,t)$ of a given configuration
$\eta \equiv (\eta_{1},\eta_{2},...,\eta_{L})$ at the time $t$ 
is given by the master
equation,
\begin{equation}
\label{master}
\frac{d}{dt} P(\eta,t)=\sum_{i}^{L}\{w_{i}(\eta^{i})P(\eta^{i},t)
-w_{i}(\eta)P(\eta,t)\},
\end{equation}
where   $\eta^{i} \equiv
(\eta_{1},\eta_{2},...,1-\eta_{i},...,\eta_{L})$.
The transition rate $w_{i}(\eta)$  for the PCCP is given by
\begin{equation}
w_{i}(\eta)=
\frac{1}{2}(1-\eta_{i})(\eta_{i-2}\eta_{i-1}+\eta_{i+1}\eta_{i+2})+\alpha
\eta_{i}.
\label{trate}
\end{equation}
If an empty site has two occupied
neighbor sites, then this site becomes occupied with a transition
rate equals to $n/2$, where $n$ is the number of the pairs of the
first neighbors occupied.
If the site is occupied, then it is becomes 
empty with a rate
$\alpha$, independently of its neighborhood.

When the lattice is entirely empty, the system is trapped in an
absorbing state.
However, for small values of the rate $\alpha$, an active state can 
be achieved
in the stationary state. In this way, a continuous phase transition
occurs in  this model with the critical point localized, according to
simulational results
\cite{mariofiore}, in $\alpha_c=0.1341(6)$. 

To develop this operator formalism, we represent the microscopic 
configurations of the lattice by the direct product of vectors
\begin{eqnarray}
\label{estado}
| \eta \rangle= \bigotimes_i | \eta_i \rangle.
\end{eqnarray}
The algebra is defined by the creation and annihilation operators 
for the site $i$
\begin{eqnarray}
\label{oper}
A_i^\dagger |\eta_i \rangle &=&(1-\eta_i)|\eta_i+1\rangle,\nonumber \\
A_i |\eta_i\rangle &=& \eta_i|\eta_i-1\rangle.
\end{eqnarray} 
In this formalism, the state of the system at time $t$ may be represent as
\begin{eqnarray}
\label{vetor}
|\psi(t)\rangle=\sum_{\{\eta\}}P(\eta,t)|\eta\rangle.
\end{eqnarray}
If we define the projection onto all possible states as
\begin{eqnarray}
\label{projec}
\langle\mbox{ }|\equiv\sum_{\{\eta\}}\langle\eta|,
\end{eqnarray}
the normalization of the state may be expressed 
as $\langle\mbox{ }|\psi\rangle=1$.

The master equation (\ref{master}) can be shown to be equivalent to
the following time evolution equation
\begin{eqnarray}
\label{eqmest}
\frac{d}{dt}|\psi(t)\rangle=S|\psi(t)\rangle,
\end{eqnarray}
where $S$ is the evolution operator, given by
\begin{equation}
S=S_{p}+\alpha S_{a},
\end{equation}
where
\begin{equation}
\label{opev}
S_p=\frac{1}{2}\sum_i\{(A_i^{\dagger}-A_iA_i^{\dagger})(A_{i-2}^{\dagger}
A_{i-2}A_{i-1}^{\dagger}A_{i-1}+A_{i+1}^{\dagger}
A_{i+1}A_{i+2}^{\dagger}A_{i+2})\}
\end{equation}
is the pair-creation operator and
\begin{equation}
S_{a}=\sum_i(A_i-A_i^{\dagger}A_i)
\end{equation}
is the annihilation operator.

\section{Supercritical series expansion}

By rescaling the time, it is possible to rewrite the evolution operator  as 
\begin{equation}
S=V+\mu W, 
\label{eqs}
\end{equation}
where
$W=S_a$, $V=2S_p$ and  $\mu=2\alpha$. 
Since the operator $W$ is associated to annihilation
process and the creation of particle is present in the operator $V$,
for small values of the parameter $\mu$ the creation process is favored,
so that the composition above is convenient for a supercritical series.
The action of each operator, in a general configuration $(\mathcal{C})$,
is explicitly shown by the expressions
\begin{equation}
\label{opsc}
W(\mathcal{C})=\sum_{i=1}^r(\mathcal{C}_i^{\prime})-r(\mathcal{C}),
\end{equation}
and
\begin{equation}
\label{opsc2}
V(\mathcal{C})=\sum_{i=1}^{p_1}(\mathcal{C}_i^{\prime\prime})
+2\sum_{i=1}^{p_2}(\mathcal{C}_i^{\prime\prime})-(p_1+2p_2)(\mathcal{C}),
\end{equation}
where $r$ is the occupied sites number, 
$(\mathcal{C}_i^{\prime})$ is obtained
replacing a particle at the site $i$ by a hole, while $p_1$ and $p_2$ are
the number of the empty sites with one and two occupied pair of neighbors, 
respectively.
Finally, $(\mathcal{C}_i^{\prime\prime})$ is the configuration obtained
replacing the hole at the site $i$ by a particle.

Since we are interested in the steady-states properties, 
it is convenient to take the
Laplace transform of the state vector,  given by
\begin{eqnarray}
\label{lt}
|\tilde{\psi}(s)\rangle=\int_{0}^{\infty}e^{-St}|\psi(t)\rangle.
\end{eqnarray}
Inserting the formal solution given by the equation (\ref{eqmest}), we find
\begin{eqnarray}
\label{tpsi}
|\tilde{\psi}(s)\rangle=(s-S)^{-1}|\psi(0)\rangle.
\end{eqnarray}
The stationary state $|\psi(\infty) \rangle \equiv \lim_{t \to \infty} |\psi(t)
\rangle$ may then be found noticing that 
\begin{equation}
|\psi(\infty) \rangle = \lim_{s \to 0} s |\tilde{\psi}(s) \rangle,
\end{equation}
which is obtained integrating  equation (\ref{lt}) by parts. 
Assuming that $|\tilde{\psi}(s) \rangle$ may be expanded in
powers of $\mu$ and using  equation (\ref{tpsi}), we have that
\begin{equation}
|\tilde{\psi}(s) \rangle = |\tilde{\psi}_0 \rangle+\mu |\tilde{\psi}_1
\rangle +\mu^2 |\tilde{\psi}_2 \rangle + \cdots = (s- V -\mu
W)^{-1} |\psi(0) \rangle. 
\end{equation}
Expanding the operator $(s- V -\mu W)^{-1}$ in powers of $\mu$, 
\begin{equation}
(s- V -\mu W)^{-1}= (s-V)^{-1} \left[ 1 + \mu (s-V)^{-1}
W 
+ \mu^2 (s-V)^{-2} W^2 + \cdots \right],
\end{equation}
and comparing each order in $\mu$, we arrive at the expressions:
\begin{equation}
|\tilde{\psi}_0 \rangle = (s-V)^{-1} |\psi(0)\rangle, 
\end{equation}
and
\begin{equation}
|\tilde{\psi}_n \rangle = (s-V)^{-1} W |\tilde{\psi}_{n-1} \rangle,
\label{rr} 
\end{equation}
for $n\ne 0$.

The action of the operator $(s-V)^{-1}$ on an arbitrary configuration
$({\mathcal{C}})$ may be found noticing that
\begin{equation}
(s-V)^{-1} ({\mathcal{C}})=s^{-1} \left\{({\mathcal{C}})+ (s-V)^{-1}V
({\mathcal{C}})\right\}, 
\end{equation}
and using the expression (\ref{opsc2}) for the action of the operator
$V$, we get
\begin{equation}
(s-V)^{-1} ({\mathcal{C}})= s_q \left\{ ({\mathcal{C}}) + (s-V)^{-1} \left[
\sum_{i=1}^{p_1} 
({\mathcal{C}}^{\prime \prime}_i)+2 \sum_{j=1}^{p_2} 
({\mathcal{C}}^{\prime \prime}_j)
\right] \right\},
\label{sv}
\end{equation}
where $s_q \equiv 1/(s+q)$ and $q=p_1+2 p_2$. 

The main obstacle to generate the standard supercritical series for this 
model is due the action of the operator $(s-V)^{-1}$ over 
a configuration $(\mathcal{P})$ without  pairs of particles,
whose result is given by
\begin{equation}
(s-V)^{-1}(\mathcal{P})=\frac{1}{s}(\mathcal{P}).
\end{equation}
The successive action of $(s-V)^{-1}$  over the configuration $(\mathcal{P})$ 
or their offsprings will  result in new
configurations whose associated coefficient 
will be proportional to $s^{-n}$ ($n>1$)  and  thus, it diverges
in the stationary state.  To overcome this difficulty, 
we  present in the next section an alternative version of PCCP model that 
avoid this divergence and, in principle, enables us to find 
asymptotic results for the critical properties of the PCCP model.

\section{Pair-creation contact process modified by a single creation
transition}

The modification consists of adding 
to the transition rate (\ref{trate})  a term of creation by single
particles similar  to the ordinary contact process
\cite{md99}, so that the transition rate is given by
\begin{equation}
w_{i}(\eta)=\frac{1}{2}(1-\eta_{i})
\{\eta_{i-2}\eta_{i-1}+\eta_{i+1}\eta_{i+2}+\gamma(\eta_{i-1}+\eta_{i+1})\}
+\alpha\eta_{i}.
\label{trate2}
\end{equation}
The operator $V$ in equation (\ref{eqs}) 
now reads $V=2(S_{p}+\gamma S_s)$, where the single
creation operator $S_s$ is given by
\begin{equation}
S_s=\frac{1}{2}\sum_i\{(A_i^{\dagger}-A_iA_i^{\dagger})(
A_{i-1}^{\dagger}A_{i-1}+A_{i+1}^{\dagger}A_{i+1})\}.
\end{equation}
In the limit of $\gamma \rightarrow 0$, we recover the pure PCCP model.

The operator  $W$ acts over $(\mathcal{C})$ according to equation 
(\ref{opsc}), whereas the action of the operator $(s-V)^{-1}$ is given by
\begin{eqnarray}
(s-V)^{-1}(\mathcal{C})& =& s_{p,\gamma}\left\{
(\mathcal{C})+(s-V)^{-1}\left[\sum_{i=1}^{p1}
(\mathcal{C}_i^{\prime\prime})+2\sum_{i=1}^{p_2}
(\mathcal{C}_i^{\prime\prime})\right]\right\}+\nonumber\\
&+&s_{p,\gamma}\gamma\left\{(s-V)^{-1}\left[\sum_{i=1}^{r_1}
(\mathcal{C}_i^{\prime\prime})+2\sum_{i=1}^{r_2}
(\mathcal{C}_i^{\prime\prime})\right]\right\},
\end{eqnarray}
where $s_{p,\gamma}=(s+[p_1+p_2]+\gamma[r_1+r_2])^{-1}$,
$p_1$  and $p_2$ (are   the number of empty sites  with one and
two pairs of occupied neighbors, respectively and  $r_1$ 
and $r_2$ are the number 
of empty sites
with one and two occupied neighbors, respectively. The expression
$(\mathcal{C}_i^{\prime\prime})$ corresponds to the configuration obtained
by replacing a hole at site $i$ by a particle.
 Differently from the pure PCCP, 
the modified PCCP  does not have 
 divergences  when one takes  the limit $s\rightarrow 0$.
 Thus, for configurations $(\mathcal{P})$, we have  that
\begin{eqnarray}
(s-V)^{-1}(\mathcal{P})=\frac{1}{s+\gamma b}\left\{
(\mathcal{P})+\sum_j(\mathcal{P}^j)\right\},
\end{eqnarray}
where $b$ is an integer number and $(\mathcal{P}^j)$ 
is a configuration originated from $(\mathcal{P})$ 
by a simple creation process.  The particular case
of the vacuum $(0)$ for which $(s-V)^{-1}(0)=1/s(0)$ does 
not diverge in the stationary state limit.

\begin{figure}[h!]
\vspace*{0.8cm}
\begin{center}
\epsfig{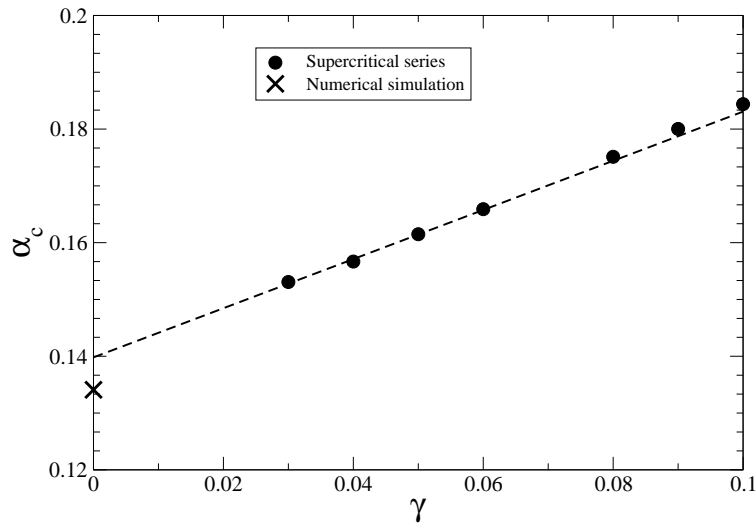}
\caption{Critical line in the plane $\alpha$ versus $\gamma$ 
for the modified
PCCP model  by supercritical series (circles). The dashed
line is a linear extrapolation toward $\gamma=0$. For comparison, 
we plotted the   numerical simulation result (cross).} 
\label{fig3}
\end{center}
\end{figure}

In the supercritical expansion the operator $(s-V)^{-1}$ acting
over any configuration (except the vacuum) generating an infinite set
of configurations, so it is impossible to  evaluate
$|{\tilde \psi(s)}\rangle$ in a closed form. We can evaluate, however,
the  ultimate survival 
probability $P_{\infty}$, which corresponds to the
coefficient of the vacuum.   We define the
series coefficients as
\begin{eqnarray}
\label{pinf}
P_{\infty}=1-\sum_i a_i(\gamma)\alpha^i.
\end{eqnarray}
Using a proper computational algorithm, we can calculate
the coefficients up to order 23, with values of $\gamma$ 
shown in  figure \ref{fig3}. The limiting factor for
this calculation is actually the memory required.
We  attempt to  obtain 
the critical properties of the PCCP model by performing the 
numerical extrapolation $\gamma\to 0$, since this approach
does not work in the case $\gamma=0$.

To analyze the series, we use the d-log Pad\'e approximants
approach. These approximants are defined as ratios of two polynomials
\begin{eqnarray}
\frac{P_L(\alpha)}{Q_M(\alpha)}=
\frac{\sum_{i=0}^{L}p_i\alpha^i}{1+\sum_{j=1}^M q_j\alpha^j}=f(\alpha).
\end{eqnarray}
In our case the function $f(\alpha)$ represents the series for 
$\frac{d}{d\alpha}\ln P_{\infty}(\alpha)$,
for a fixed value for the parameter $\gamma$. Therefore, we are able to
obtain approximants satisfying the condition
$L+M\leq 22$. One verifies that  diagonal ($L=M$) and
near-diagonal approximants usually exhibit better convergence
properties, so that  we will 
restrict our calculations to the set of approximants
such that $L=M+\theta$, with $\theta=0,\pm 1$.

In the neighborhood of the critical point the ultimate survival
probability  behaves as 
$P_{\infty}\sim(\alpha_c-\alpha)^{\beta}$.   
 The critical point $\alpha_c$ is determined 
by the pole of the approximant $F_{LM}(\alpha)$ and the exponent 
$\beta$ by the residue associated with this pole.  

In figure  \ref{fig3} the estimates for critical points $\alpha_c$ 
are depicted for different values of the parameter $\gamma$. 
Using a linear extrapolation of these data, we obtain
the estimation for the asymptotic limit $\gamma \rightarrow 0$ as
$\alpha_c=0.1398(7)$, in
disagreement 
with  the simulational result $\alpha_c=0.1341(6)$ \cite{mariofiore}.
We remark  that decreasing the parameter $\gamma$, the dispersion
of the approximants increases and the precision of the result becomes smaller.

\begin{figure}[h!]
\vspace*{0.8cm}
\begin{center}
\epsfig{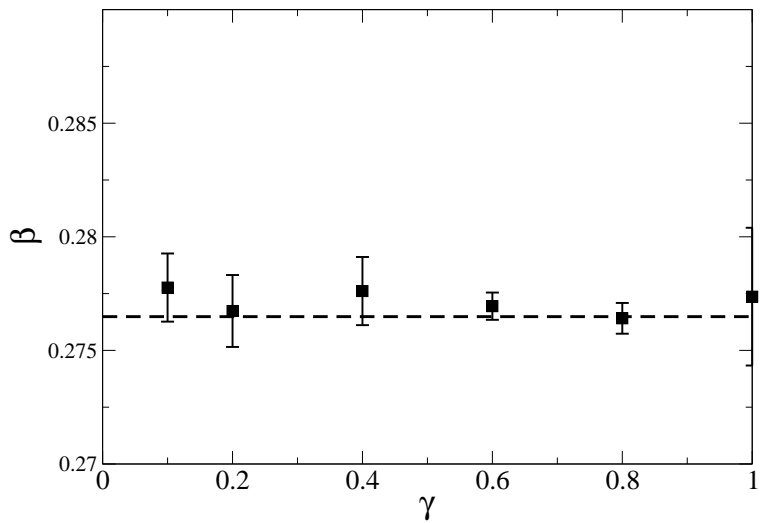}
\caption{Exponent $\beta$ for the modified PCCP model
as a function of $\gamma$.}
\label{figbeta}
\end{center}
\end{figure} 

The exponent $\beta$, as shown in figure \ref{figbeta}, 
is in accordance with the DP universality class value
for any value of the parameter $\gamma$,  as expected.

However, the uncertainties on the location 
of the critical point  put in doubt the accuracy of the
supercritical series around $\gamma \approx 0$.  
In fact, what happens is that the coefficients of the configurations
 $(\mathcal{P})$ in the steady-state behaves as 
\begin{eqnarray}
(s-V)^{-1}(\mathcal{P})\sim\frac{1}{\gamma b}(\mathcal{P}),
\end{eqnarray}
diverging as $\gamma\to 0$. This    behavior 
causes a ill-conditioned series for $\gamma\approx 0$, becoming impossible
a more accurate result for the critical point in this limit.

\section{Subcritical series expansion}

To develop subcritical series expansion, 
we rewrite the evolution operator as 
\begin{equation}
S=W+\lambda (S_{p}+\gamma S_{s}), 
\end{equation}
where $\lambda=\alpha^{-1}$. 
Since the operator $W$ is associated to annihilation
process and, for small values of the parameter $\lambda$,
this annihilation process is favored, this expansion describes 
indeed the subcritical regime.
Differently from the supercritical case, the series here are generated
directly, without the necessity of taking the Laplace transform.

The operator $W$, associated to  
annihilation particles process, is expressed as $W=\sum_i W_i$, 
with $W_i=A_i-A_i^{\dagger}A_i$. Each term  $W_i$ has the following set
of right and left eigenvectors
\begin{equation}
|0\rangle \equiv |\circ \rangle , \hspace{1 cm}\langle  0| 
\equiv \langle \circ| + \langle \bullet|,
\end{equation}
with eigenvalue $\Lambda_{0}=0$ and
\begin{equation}
|1\rangle \equiv -|\circ \rangle + |\bullet \rangle, 
\hspace{1 cm}\langle  1| \equiv \langle \bullet|, 
\end{equation}
with eigenvalue $\Lambda_{1}=-1$. 

To  find the steady state vector $|\psi\rangle$, that satisfies
the steady condition
$S|\psi\rangle=0$, we assume that
\begin{equation}
\label{perturbative}
|\psi\rangle=|\psi_{0}\rangle 
+\sum_{\ell=1}^{\infty}\lambda^{\ell}|\psi_{\ell}\rangle,
\end{equation}
where $|\psi_{0}\rangle $  is the steady solution
of the non-interacting term $W$ satisfying the stationary condition
\begin{equation}
W|\psi_{0}\rangle=0.
\end{equation}
The vectors $|\psi_{\ell}\rangle$ can be generated recursively
from the  initial state $|\psi_{0}\rangle$. 
Following Dickman \cite{dick89}, we get the recursion relation
\begin{equation}
\label{recursive}
|\psi_{\ell}\rangle=-R(S_{p}+\gamma S_{s})|\psi_{\ell-1}\rangle.
\end{equation}
The operator $R$ is the inverse of $W$ in the subspace of
vectors with eigenvalues and is given by
\begin{equation}
\label{R}
R=\sum_{n(\neq 0)}|\phi_{n}\rangle \frac{1}{\Lambda_{n}}\langle  \phi_{n}|,
\end{equation}
where $|\phi_{n}\rangle$ and $\langle  \phi_{n}|$  are 
right and left eigenvectors of $W$,
respectively, with nonzero eigenvalue $\Lambda_{n}$. 
Since the creation of particles is catalytic,  if we start from 
steady state of the noninteracting term, that corresponds to the vacuum 
state, we will obtain a trivial steady vector.
To overcome this problem, it is necessary to introduce a modification
on the rules of the model. The necessity of changing the initial state
in  systems with absorbing states in order
to get nontrivial steady states 
have been considered previously by Jensen and Dickman 
\cite{dj91,jd94} 
and more recently by de Oliveira \cite{mj06}. 

The  modification we have performed \cite{mj06} consists in introducing 
a spontaneous creation of particles.
For the case $\gamma=0$ (pure PCCP) the creation
occurs  in two  adjacent sites,  chosen to be  $i=0$ and
$i=1$. This modification leads to the following expression to the
 operator $W$
\begin{equation} 
W=\sum_{i} W_i+ q(U_{0}+U_{1}-W_0-W_1),
\label{wo}
\end{equation}
where $q$ is supposed to be a small parameter and 
$U_i=A_i^{\dagger}-A_iA_i^{\dagger}$.
The steady state $|\psi_{0}\rangle$ of 
$W$ is not
the vacuum state anymore.  Now, it is given by
\begin{equation}
|\psi_{0}\rangle=|.0.\rangle+
2q|.10.\rangle+q^{2}|.11.\rangle,
\end{equation}
where all sites before and after the symbol ``.'' are empty.

Two remarks are in order. First, only the last term in
$|\psi_{0}\rangle$ will give nonzero contributions to the expansion
so that $|\psi_\ell \rangle$, $\ell\geq 1$, will be of the
order $q^2$. Second, 
Although the change in $W$ will cause a
change in $R$, only the terms of zero order in the expansion in $q$, 
given by the right-hand side of equation (\ref{R}), will be necessary
since the corrections in $R$ will contribute to terms of order larger than
$q^2$. 
For instance, the two first vectors, $|\psi_{1} \rangle$ and $|\psi_{2}
\rangle$, for the pure PCCP are given by
\begin{equation}
\label{vector1}
|\psi_{1}\rangle=q^{2}\{2|.1.\rangle+
|.11.\rangle+
|.101.\rangle+
\frac{2}{3}|.111.\rangle \},
\end{equation}
and
\[
|\psi_{2}\rangle=q^{2}\{\frac{2}{3}|.1.\rangle
+\frac{1}{3}|.11.\rangle
+\frac{1}{3}|.101.\rangle
+\frac{2}{9}|.111.\rangle+
\]
\begin{equation}
+\frac{2}{3}|.1001.\rangle
+\frac{4}{9}|.1101.\rangle
+\frac{4}{9}|.1011.\rangle
+\frac{1}{3}|.1111.\rangle\},
\end{equation}
 where the translational invariance of the system is assumed.

For $\gamma \neq 0$, it suffices to consider the spontaneous creation
in just one site. This modifications leads to the following expression
\begin{equation} 
W=\sum_{i} W_i+ q(U_{0}-W_0),
\end{equation}
instead of equation (\ref{wo}).

From the series expansion of the state vector $|\psi \rangle$, it is possible
to determine several quantities.
In this paper, we will be concerned only with the series 
expansion for the total number of particles $N$, given by
\begin{equation}
N=\langle .0.|\sum_{i}n_{i}| \psi \rangle.
\end{equation}
One can show that the coefficient of
$\lambda^\ell$ in the expansion for $N$ is simply the coefficient of
$|.1.\rangle$ in $|\psi_\ell\rangle$. This allows us
to get a longer series for the number of particles. For both modified
and pure PCCP,  we have obtained a series with 38 terms. 

The critical behavior of $N$ obeys the following relation
\cite{mj06}
\begin{equation}
N \sim (\alpha_c -\alpha)^{-\nu_{\,||}(1+\eta)},
\end{equation}
where $\nu_{\,||}$ and $\eta$ are the exponents related to the 
time correlation length and to the growth
of the number of particles, respectively.
Through an analysis of the d-log Pad\'e approximants, 
like that performed for supercritical series expansion, 
we are able to calculate the critical point and its associated 
exponent. Using the series for the number of particles $N$ the values
obtained were $\alpha_c=0.1310(14)$ and 
$\nu_{\parallel}(1+\eta)=2.68(15)$. These values do not
agree with the best estimates, $\alpha_c=0.1341(6)$ \cite{mariofiore}
and $\nu_{\parallel}(1+\eta)=2.2777305$ \cite{j99}. Another estimate
is obtained by performing a non-homogeneous d-log Pad\'e approximant 
\cite{far94}.
Using this approach we obtained the value
$\alpha_c=0.1337(17)$, closer
to the value obtained by numerical simulations \cite{mariofiore}.

More reliable estimates are obtained when we
determine the  Pad\'e approximants for the series
$(\alpha_c-\alpha)(d/d \alpha)$ ${\rm ln} N=\theta$
\cite{dj91,wgdstilckcross}, called  biased Pad\'e approximants.  
Considering a given Pad\'e approximant $[L/M]$  and
a trial value  of $\tilde \alpha_{c}$, we develop the series above obtaining
$\theta(\tilde\alpha_{c})$.
We can build curves for different Pad\'e approximants by repeating this
procedure for several trials $\tilde \alpha_{c}$
and we expect that they intercept at  the critical  
point ($ \alpha_{c}$, $ \theta(\alpha_{c})$).

\begin{figure}
\vspace{1.5cm}
\begin{center}
\epsfig{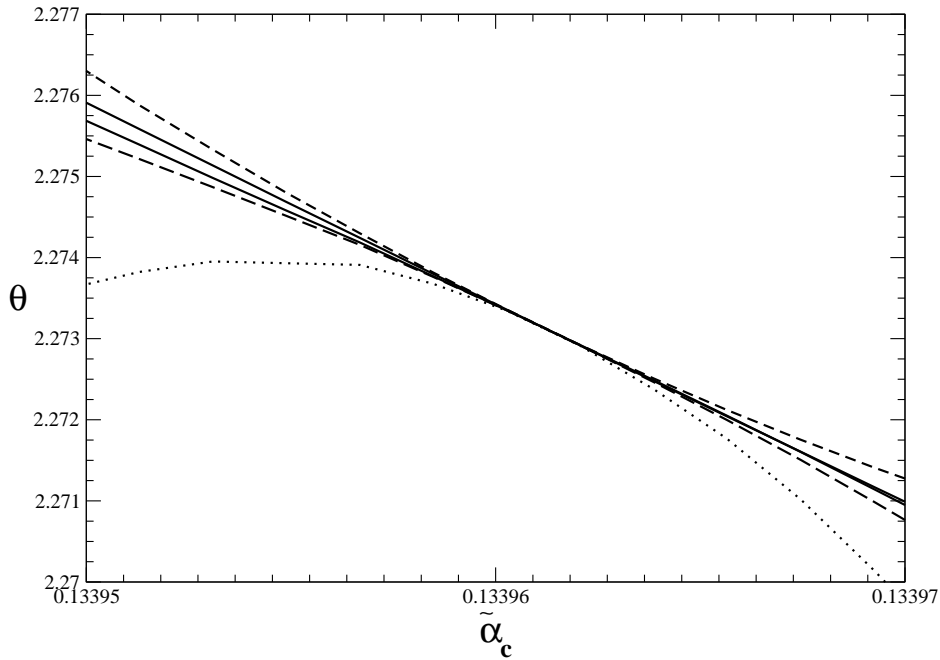}
\vspace{0.1 cm}
\caption{Biased estimates of $ \theta=\nu_{\,||}(1+\eta)$ as
a function of $\tilde{\alpha_{c}}$ derived from Pad\'e approximants
to the series $(\tilde{\alpha_c}-\alpha)(d/d \alpha)$ 
$\ln N=\theta$ evaluated at $\tilde\alpha_{c}$ for the PCCP.
The approximants shown are [17/18], [18/17], [18/18], [18/19], and [19/18].}
\label{padepccp}
\end{center}
\end{figure}

From the figure \ref{padepccp}
we see a very narrow intersection
of the Pad\'e approximants, revealing the utility of this approach.
However, as pointed out by Guttmann \cite{guttmann}, it is difficult to
estimate uncertainties in series calculations. Thus,
in order to give a more  realistic estimate of the 
quantities measured here and their associated uncertainties, 
we have estimated them by taking into account the first and last
crossings among various Pad\'e approximants. 
The estimate of $\alpha_c$ for the PCCP, $\alpha_c=0.13396(1)$, 
 in excellent agreement with the corresponding value 
obtained from recent numerical simulations \cite{mariofiore}.
Finally, the exponent obtained is $\nu_{\parallel}(1+\eta)=2.274(3)$,
in  agreement with the best estimation of $\nu_{\parallel}(1+\eta)$
for the DP universality class \cite{j99}.  

\begin{figure}
\vspace{1.5cm}
\begin{center}
\epsfig{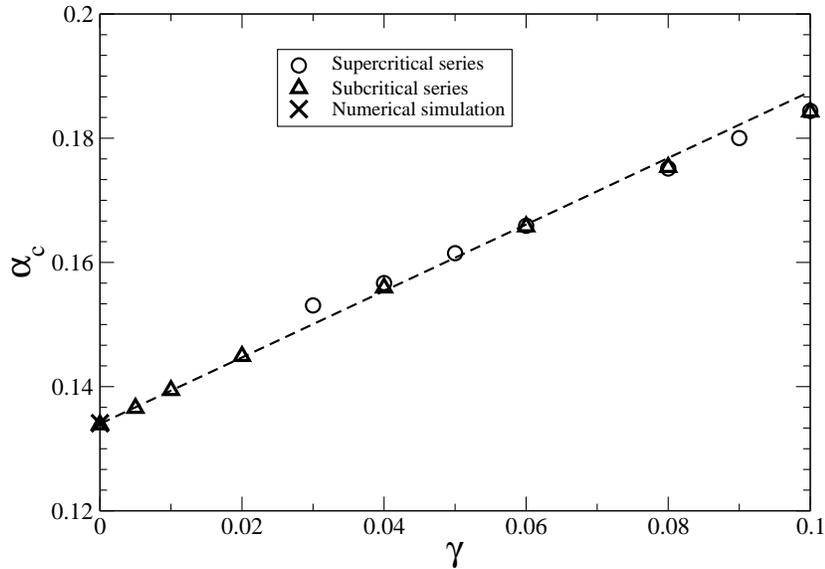}
\vspace{0.1 cm}
\caption{Comparative results for the modified PCCP model using 
supercritical (circles), subcritical (triangles) series 
and numerical simulations (cross).}
\label{fig5}
\end{center}
\end{figure}

To stress the difference between  subcritical and supercritical
approaches, we have also developed subcritical
series expansion for the modified  PCCP model
for some values of $\gamma$, as 
shown in figure \ref{fig5}. In the limit $\gamma \rightarrow 0$,
the subcritical approach gives $\alpha_c=0.1340(1)$, 
which is in good agreement 
with the estimate $\alpha_c=0.13396(1)$ for $\gamma=0$ and 
$\alpha_c=0.1341(6)$ obtained from numerical simulations \cite{mariofiore}. 
On the other hand,
the extrapolation for $\gamma \rightarrow 0$ in the supercritical
case does not lead to the correct value, 
as can be  seen in  figure \ref{fig3}. 

\section{Conclusions}

In this work, we have considered the pair-creation
contact  process by carrying out a 
comparative study between subcritical and supercritical
expansions. Differently from
the contact process, the supercritical expansion
in its canonical formulation \cite{dj91} is intrinsically 
ill-conditioned for the PCCP. In fact, divergences 
in coefficients of the supercritical series occur 
for any model with creation by cluster of particles.  
As an attempt to  circumvent the divergences, we introduced
and analyzed a modified version of the PCCP. However  this procedure
does not lead to the correct result when one recovers the original model.
It would be interesting to search an alternative procedure 
to develop supercritical series for such models.  
We have shown here that subcritical series expansion 
provides results comparable
with recent numerical simulations \cite{mariofiore}.

A natural extension of this work is the use of 
subcritical series to study models with creation by 
clusters (pairs and triplets) and diffusion of particles 
\cite{dicktom91,mariofiore}. Results from numerical 
simulations \cite{mariofiore,fontanari} suggest that these models 
display a tricritical point. Thus this approach, associated 
with an analysis by means of partial differential 
approximants in two-variables, could be used to determine the existence 
of a tricritical point.  
In fact, such an approach has already been shown 
to be useful for the location 
of a multicritical point in a generalized contact process 
\cite{wgdstilckcross}.

\vspace*{0.5cm}

\section*{Acknowledgement}

C. E. Fiore and W.G. Dantas thank the financial 
support from  Funda\c{c}\~ao de Amparo \`a Pesquisa do
Estado de S\~ao Paulo (FAPESP) under Grants No. 05/04459-1 and
06/51286-8.

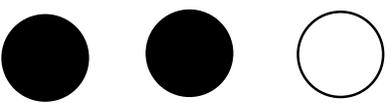
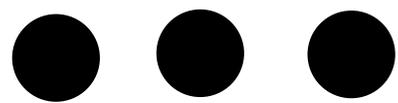
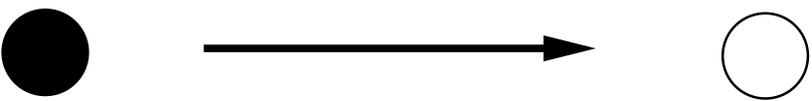
\end{document}